\documentclass{emulateapj}

\shorttitle{Turbulent Shear Acceleration}

\shortauthors{Yutaka Ohira}

\begin{document}

\title{Turbulent Shear Acceleration}

\author{Yutaka Ohira}

\begin{abstract}
We consider particle acceleration by large-scale incompressible turbulence 
with a lengthscale larger than the particle mean free path.
We derive an ensemble-averaged transport equation of energetic charged  
particles from an extended transport equation 
which contains the shear acceleration.
The ensemble-averaged transport equation describes 
particle acceleration by incompressible turbulence (turbulent shear acceleration).
We find that for Kolmogorov turbulence, 
the turbulent shear acceleration becomes important in small scale.
Moreover, by Monte Carlo simulations, we confirm that the 
ensemble-averaged transport equation describes the turbulent shear acceleration. 
\end{abstract}


\affil{Department of Physics and Mathematics, Aoyama Gakuin University, 
5-10-1 Fuchinobe, Sagamihara 252-5258, Japan; ohira@phys.aoyama.ac.jp}
%%%%%%%%%%%%%%%%%%%%%%%%%%%%%%%%%%%%%%%%%%%%%%%%%%%%%%%%%%%%%%%%%%%%%%
\section{Introduction}
Charged particles are accelerated to relativistic energies 
in many astrophysical objects. In addition, turbulence is also expected. 
In fact, strong turbulence is observed in recent two and three 
dimensional simulations for supernova remnants (SNRs) 
\citep{giacalone07,inoue09,guo12,caprioli12}, 
pulsar wind nebulae (PWNe) \citep{komissarov04,del04,porth12}, 
astrophysical jets \citep{aloy99,mizuta10,lopez12}, etc. 

There are mainly two acceleration mechanisms by turbulence.
One is due to wave-particle interactions, where the particle mean free path 
is comparable to the wavelength of electromagnetic 
fluctuations \citep[e.g][]{skilling75,schlickeiser98}.
The other is due to large-scale fluctuations of plasma flows, where the particle mean 
free path is smaller than turbulent scales \citep[e.g.][]{bykov93}. 
Turbulence is generally divided by compressible and incompressible modes. 
Particle acceleration by large-scale compressible turbulence has been 
discussed in many paper \citep[e.g.][]{bykov82,ptuskin88,jokipii10}.
However, particle acceleration by large-scale incompressible turbulence 
(turbulent shear acceleration) has not been investigated in detail, while 
\citet{bykov83} has briefly discussed the turbulent shear acceleration. 

Particle acceleration by a simple incompressible flow (shear flow) 
has already investigated in many paper 
\citep[e.g.][]{berezhko81,earl88,webb89,ostrowski90,rieger06}.
However, shear flows are potentially unstable to the Kelvin-Helmholtz instability  
and produce turbulence. 
Therefore, the turbulent shear acceleration is expected to be important.
In this Letter, we investigate the turbulent shear acceleration 
by considering ensemble average of an extended transport equation 
which includes particle acceleration by shear flows.

We first derive an ensemble-averaged transport equation in Section 2, 
and provide its analytical solutions for simple cases in Section 3. 
We then perform Monte Carlo simulations in Section 4.
Section 5 is devoted to the discussion. 
%%%%%%%%%%%%%%%%%%%%%%%%%%%%%%%%%%%%%%%%%%%%%%%%%%%%%%%%%%%%%%%%%%%%%%
\section{Derivation of the Ensemble-Averaged Transport Equation}
\label{sec:2} 
In this section, we derive the ensemble-averaged transport equation 
of energetic particles. 
Propagation and acceleration of energetic charged particles in a plasma flow
are described by a transport equation.
\citet{parker65} derived the transport equation which includes spatial diffusion, 
convection, and adiabatic acceleration. 
After that his work was extended by several authors.
For isotropic diffusion and a nonrelativistic plasma flow, 
the extended transport equation is given by 
(Equation (4.5) of \citet{webb89} and Equation (9) of \citet{williams93})
\begin{eqnarray}
\frac{\partial F}{\partial t} 
&+& U_i \frac{\partial F}{\partial x_i} 
- \frac{\partial}{\partial x_i} \left ( \kappa \frac{\partial F}{\partial x_i} \right ) 
- \frac{p}{3}\frac{\partial U_i}{\partial x_i}\frac{\partial F}{\partial p} \nonumber \\
&-&\frac{1}{p^2}\frac{\partial }{\partial p} \left( \kappa \Gamma \frac{p^4}{v^2} \frac{\partial F}{\partial p}\right) \nonumber \\
&-& \frac{1}{p^2}\frac{\partial }{\partial p} \left( \kappa  \frac{DU_i}{Dt}\frac{DU_i}{Dt} \frac{p^4}{v^4} \frac{\partial F}{\partial p}\right) \nonumber \\
&+&\frac{\partial }{\partial x_i} \left( \kappa  \frac{DU_i}{Dt}\frac{p}{v^2} \frac{\partial F}{\partial p}\right)\nonumber \\
&+&\frac{1}{p^2}\frac{\partial }{\partial p} \left( \kappa  \frac{DU_i}{Dt}\frac{p^3}{v^2} \frac{\partial F}{\partial x_i}\right)=0~~,
\label{eq:transporteq}
\end{eqnarray}
where $\Gamma$ is defined by
\begin{equation}
\Gamma = \frac{1}{5}\left( \frac{\partial U_i}{\partial x_j}\frac{\partial U_j}{\partial x_i} + \frac{\partial U_i}{\partial x_j}\frac{\partial U_i}{\partial x_j}  \right) -\frac{2}{15}\frac{\partial U_i}{\partial x_i}\frac{\partial U_j}{\partial x_j} ~~.
\end{equation}
$F(p,{\bf x},t), x_i, U_i, \kappa(p)$ are the distribution function, 
position, plasma velocity, and spatial diffusion coefficient, respectively. 
$v$ and $p$ are the particle velocity and four momentum 
in the fluid rest frame, respectively. 
The spatial diffusion coefficient, $\kappa$, is represented by
$\kappa(p)=\tau(p)v^2/3$ for isotropic diffusion, where $\tau(p)$ is 
the mean scattering time and $\tau v$ is the particle mean free path. 
The first four terms of Equation (\ref{eq:transporteq}) are the same as the Parker equation
and the others are additional terms. 
The fifth term describes the shear acceleration and the sixth term becomes 
important for $v\sim U_i$.

In order to understand essential features of the turbulent shear acceleration, 
we consider incompressible turbulence ($\partial U_i/\partial x_i=0$) and 
do not take into account the spatial transport, that is, 
we consider the spatially averaged distribution function, $V^{-1}\int_{V} F d^3x$, where $V$ is 
the system volume that we consider. 
By integrating Equation (\ref{eq:transporteq}), the extended transport equation can be reduced to 
\begin{eqnarray}
\frac{\partial }{\partial t} \frac{1}{V}\int_V F d^3x
&-&\frac{1}{p^2}\frac{\partial }{\partial p} \left ( \frac{\tau p^4}{3} \frac{\partial}{\partial p}
 \frac{1}{V}\int_{V} \Gamma F  d^3x \right ) \nonumber \\
&-&\frac{1}{p^2}\frac{\partial }{\partial p} \left ( \frac{\tau p^4}{3v^2}\frac{\partial}{\partial p}
\frac{1}{V}\int_{V} \frac{DU_i}{Dt}\frac{DU_i}{Dt} F d^3x \right )  \nonumber \\
&+& Q(p) =0~~,
\label{eq:req}
\end{eqnarray}
where $Q(p)$ is the particle flux passing through the surface of the integrated volume. 
As long as we consider a timescale smaller than $V^{1/3}/U_i$ and $V^{2/3}/\kappa$, 
we can neglect the particle flux, $Q(p)$.
In other words, we can neglect escape of particles from the system 
when we consider a sufficiently large system size.

In this Letter, we assume that the plasma velocity field, $U_i({\bf x},t)$, 
is static, random, statistically homogenous and isotropic incompressible turbulence, 
that is, $U_i = \delta u_i({\bf x})$ and $\langle \delta u_i \rangle =0$, 
where $\langle ...\rangle$ denotes ensemble average. 
The correlation function of the plasma velocity field is given by
\begin{equation}
\langle \delta u_i({\bf x})\delta u_j({\bf x'})\rangle=\int \frac{{\rm d}^3k}{(2\pi)^3}
K_{ij}({\bf k})e^{i\{ k_l(x_l-x'_l)}~~,
\label{eq:coreelation}
\end{equation}
and 
\begin{equation}
K_{ij}({\bf k}) = S(k)\left( \delta_{ij} - \frac{k_i k_j}{k^2}  \right)~~,
\label{eq:vspectrum}
\end{equation}
where $k$ and $S(k)$ are the wavenumber and 
spectrum of incompressible turbulence, respectively. 
As long as we consider only particle acceleration, 
we can assume the velocity field to be static 
when the scattering timescale, $\tau$, is smaller than 
the variable timescale of fluid, $T \sim (k \times \max \left\{\delta u,v_{\rm ph}\right\})^{-1}$,
where $ v_{\rm ph} $ is a phase velocity. 
In this letter, we consider $\tau v k < 1$ and $v>\max \left\{\delta u,v_{\rm ph}\right\}$, 
so that $\tau / T \sim \tau k\times \max \left\{\delta u,v_{\rm ph}\right\} <\tau vk< 1$.
Hence, we can assume a static velocity field in this letter.

The distribution function of particles can also be divided by an 
ensemble-averaged component and a fluctuated one, that is, 
$F=f+\delta f$ and $\langle F\rangle=f$.
The spatial average in Equation (\ref{eq:req}) can be interpreted as ensemble 
average because we consider a system size larger than the turbulent scale.
Then, from Equation (\ref{eq:req}), the ensemble-averaged transport equation is represented by 
\begin{equation}
\frac{\partial f}{\partial t}  - \frac{1}{p^2}\frac{\partial }{\partial p} \left\{ \frac{\tau p^4}{3} \left ( \langle \Gamma \rangle  + \left \langle u_j\frac{\partial \delta u_i}{\partial x_j}u_l\frac{\partial \delta u_i}{\partial x_l} \right \rangle \frac{1}{v^2} \right ) \frac{\partial f}{\partial p}\right\}=0~~,
\label{eq:aat}
\end{equation}
where we have assumed that distributions of $\delta f$ and $\delta u_i$ are symmetric 
about the mean values, $f$ and $0$, respectively, so that third moments are zero. 
From Equations (\ref{eq:coreelation}), (\ref{eq:vspectrum}), and (\ref{eq:aat}), 
the ensemble-averaged transport equation can be represented by
\begin{equation}
\frac{\partial f}{\partial t} - \frac{1}{p^2}\frac{\partial }{\partial p} \left( p^2D_{\rm TSA}  \frac{\partial f}{\partial p}\right)=0~~,
\label{eq:f}
\end{equation}
where the momentum diffusion coefficient, $D_{\rm TSA}(p)$, is given by 
\begin{equation}
D_{\rm TSA}(p) = \frac{2}{9}p^2\tau(p) \int \frac{{\rm d}^3k}{(2\pi)^3}
 S(k) k^2 \left ( \frac{3}{5}+ \frac{\langle \delta u^2\rangle}{v^2} \right )~~.
\label{eq:ds}
\end{equation}
We here consider turbulence with a large lengthscale compared with the particle 
mean free path, $\tau v$, so that the upper limit of $k$-integral 
should be limited by $\min \{k_{\max},k_{\rm res}\}$ where $k_{\max}$ 
is the maximum wavenumber of turbulence and 
$k_{\rm res}\approx (\tau v)^{-1}$. 
The momentum diffusion coefficient, $D_{\rm TSA}$, is dominated by small scale turbulence 
when $k^5S(k)$ is an increasing function of $k$. 
Therefore, the turbulent shear acceleration becomes important 
in the small scale for a Kolmogorov-like spectrum 
($S(k)\propto k^{-11/3}$).

%%%%%%%%%%%%%%%%%%%%%%%%%%%%%%%%%%%%%%%%%%%%%%%%%%%%%%%%%%%%%%%%%%%%%%
\section{Analytical solution}
\label{sec:3}
In this section, we present specific expressions of the momentum diffusion 
coefficient and analytical solutions of the ensemble-averaged transport equation 
for simple velocity spectra.
We especially focus on the turbulent shear acceleration of 
relativistic particles ($v\approx c$) in nonrelativistic turbulence 
($\langle \delta u^2\rangle \ll c^2$), 
so that we neglect the term of $\langle \delta u^2\rangle/v^2$ in 
Equation~(\ref{eq:ds}).
We here assume a functional form of the mean scattering time, $\tau(p)$, 
to be $\tau_0 (p/p_0)^{\alpha}$, where $p_0$ and $\tau_0$ are 
the initial four momentum and the mean scattering time of particles 
with $p_0$, respectively. 
To make the expression simple, hereafter the four momentum, time, 
and momentum diffusion coefficient are normalized by $p_0, \tau_0$, 
and $p_0^2\tau_0^{-1}$, respectively.
Normalized quantities are denoted with a tilde.

For a static monochromatic spectrum of incompressible turbulence, 
$S(k)$ is given by
\begin{equation}
S(k)= \frac{\langle \delta u^2\rangle (2\pi)^2}{4k_0^2} \delta(k-k_0)~~.
\label{eq:sm}
\end{equation}
From Equations (\ref{eq:ds}) and (\ref{eq:sm}), the momentum diffusion coefficient is represented by
\begin{equation}
\tilde{D}_{\rm TSA} = \frac{\left(\tau_0 v k_0\right)^2 }{15}\frac{\langle\delta u^2\rangle}{v^2} \tilde{p}^{2+\alpha}~~.
\label{eq:dm}
\end{equation}

For a static Kolmogorov-like spectrum of incompressible turbulence, 
we assume that $S(k)$ is given by
\begin{equation}
S(k)= \frac{\langle\delta u^2\rangle (2\pi)^2}{6\left(k_0^{-2/3}-k_{\max}^{-2/3}\right)} k^{-11/3} ~( {\rm for}~k_0\leq k\leq k_{\max})
\label{eq:sk}
\end{equation}
Then, from Equations (\ref{eq:ds}) and (\ref{eq:sk}), the momentum diffusion coefficient is represented by
\begin{equation}
\tilde{D}_{\rm TSA} \approx \frac{\left(\tau_0 vk_0\right)^2}{30}  \frac{\langle\delta u^2\rangle}{v^2}   \tilde{p}^{2+\alpha}\left(\frac{\min\{k_{\max},k_{\rm res}\}}{k_0}\right)^{4/3}~~,
\label{eq:dturbsu}
\end{equation}
where we have assumed $k_0\ll \min\{k_{\max},k_{\rm res}\}$. 
The factor, $\left(\min\{k_{\max},k_{\rm res}\}/k_0\right)^{4/3}$, 
is expected to be large. 
Therefore, the Kolmogorov-like turbulent cascade enhances the turbulent shear 
acceleration.
For $k_{\rm res}<k_{\max}$, $\tilde{D}_{\rm TSA}$ is represented by
\begin{equation}
\tilde{D}_{\rm TSA} \approx \frac{\left(\tau_0 vk_0\right)^{2/3}}{30}  \frac{\langle\delta u^2\rangle}{v^2}   \tilde{p}^{2-\alpha/3}~~.
\label{eq:dturbs}
\end{equation}
Therefore, the momentum diffusion coefficient can be represented by 
$\tilde{D}_{\rm TSA}=\tilde{D}_0 \tilde{p}^{2+\beta}$ for above simple cases, 
where $\beta=\alpha$ for the monochromatic spectrum 
and the Kolmogorov spectrum of the case $k_{\rm res}>k_{\max}$, 
and $\beta=-\alpha/3$ for the Kolmogorov spectrum of the case 
$k_{\rm res}<k_{\max}$.

We next discuss analytical solutions of the ensemble-averaged transport equation. 
We assume that particles are uniformly distributed in the three dimensional space 
and injected at time, $\tilde{t}=0$, with the four momentum, $\tilde{p}=1$. 
Then, the ensemble-averaged transport equation is represented by 
\begin{equation}
\frac{\partial f}{\partial \tilde{t}} - \frac{1}{\tilde{p}^2}\frac{\partial }{\partial \tilde{p}} \left( \tilde{p}^2\tilde{D}_{\rm TSA}\frac{\partial f}{\partial \tilde{p}}\right)=\frac{N}{4\pi}\delta(\tilde{t})\delta(\tilde{p}-1)~~,
\end{equation}
where $N$ is the number of injected particles.
If the momentum diffusion coefficient is represented by 
$\tilde{D}_{\rm TSA}=\tilde{D}_0 \tilde{p}^{2+\beta}$, for $\beta \neq 0$, 
the solution is given by \citep{berezhko82,rieger06}
\begin{eqnarray}
f(\tilde{p},\tilde{t}) &=& \frac{N}{4\pi |\beta|\tilde{D}_0 \tilde{t}}\tilde{p}^{-(3+\beta)/2}
\exp{\left(-\frac{1+\tilde{p}^{\beta}}{\beta^2\tilde{D}_0\tilde{t}}\right)} \nonumber \\
&&\times I_{|1+3/\beta|}\left[\frac{\tilde{p}^{-\beta/2}}{\beta^2\tilde{D}_0\tilde{t}}\right]~~,
\label{eq:analytical}
\end{eqnarray}
where $I_{\nu}$ is the modified Bessel function of the first kind. 
The solution approches $\tilde{p}^3f\propto \tilde{p}^{-\beta}$ for $ \tilde{p} > 1$.
For $\beta=0$, the solution is given by \citep{rieger06}
\begin{equation}
f(\tilde{p},\tilde{t}) = \frac{N}{(4\pi)^{3/2} \sqrt{\tilde{D}_0 \tilde{t}}}
\exp{\left\{-\frac{ \left(\ln\tilde{p}+3\tilde{D}_0\tilde{t} \right)^2 }{4\tilde{D}_0\tilde{t}} \right\}} ~~,
\label{eq:analytical0}
\end{equation}
and the evolution of the mean momentum, 
$\tilde{p}_{\rm m}(\tilde{t})=N^{-1}\int \tilde{p}f(\tilde{p},\tilde{t})4\pi \tilde{p}^2{\rm d}\tilde{p}$,
is given by
\begin{equation}
\tilde{p}_{\rm m}(\tilde{t}) = \exp{\left( 4\tilde{D}_0\tilde{t}\right)}~~.
\label{eq:mean}
\end{equation}
Note that solutions of Equations (\ref{eq:analytical}) and (\ref{eq:analytical0}) 
are not valid for $\tilde{t}\ll1$ and $\tilde{p}\gg1$ because of causality.

%%%%%%%%%%%%%%%%%%%%%%%%%%%%%%%%%%%%%%%%%%%%%%%%%%%%%%%%%%%%%%%%%%%%%%
\section{Monte Carlo Simulation}
\label{sec:4}
\begin{figure}
\plotone{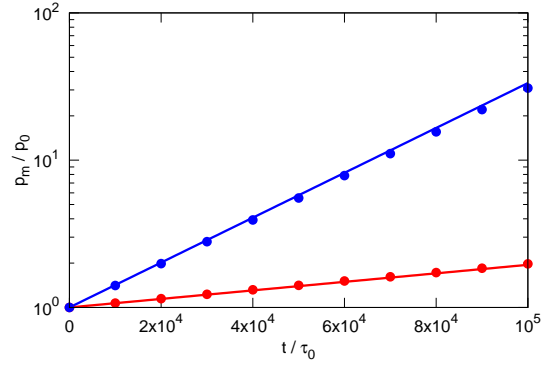} 
\caption{Time evolution of the mean four momentum for $\alpha=0$ and 
$\tau_0 c k_0 = 10^{-1}$. 
The dots and solid lines show the results of Monte Carlo simulations 
and analytical solutions ( Equations (\ref{eq:dm}), (\ref{eq:dturbsu}) 
and (\ref{eq:mean})), respectively.
The red and blue show cases of the monochromatic and Kolmogorov spectra 
with $\tau_0ck_{\max}=10^{-1/3}$, respectively.
\label{fig:1}}
\end{figure}
\begin{figure}
\plotone{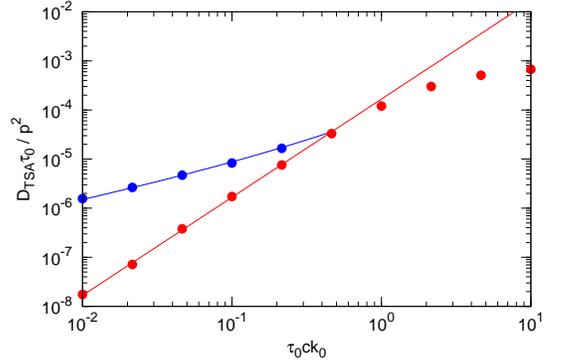} 
\caption{Wavenumber dependence of the momentum diffusion coefficient 
for $\alpha=0$. 
The dots and solid lines show the results of Monte Carlo simulations 
and analytical solutions of Equations (\ref{eq:dm}) and (\ref{eq:dturbsu}), 
respectively.
The red and blue show cases of the monochromatic and Kolmogorov spectra 
with $\tau_0ck_{\max}=10^{-1/3}$, respectively.
\label{fig:2}}
\end{figure}
\begin{figure}
\plotone{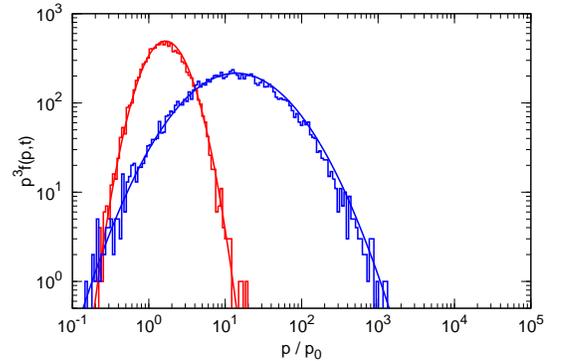} 
\caption{Distribution function at $t/\tau_0=10^5$ for $\alpha=0$ 
and $\tau_0 c k_0 = 10^{-1}$.
The histograms and solid lines show the results of Monte Carlo simulations 
and analytical solutions of Equations (\ref{eq:analytical0}), respectively.
The red and blue show cases of the monochromatic and Kolmogorov spectra 
with $\tau_0ck_{\max}=10^{-1/3}$, respectively.
\label{fig:3}}
\end{figure}
\begin{figure}
\plotone{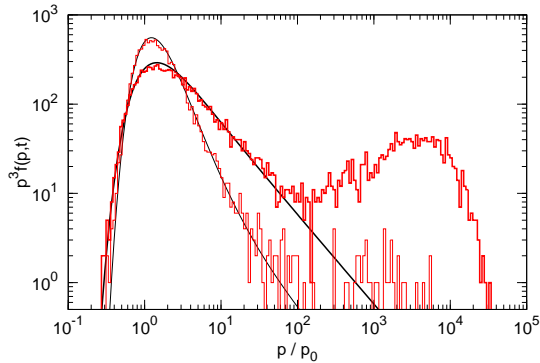} 
\caption{Distribution function for the monochromatic spectrum with 
$\tau_0 c k_0 = 0.01$ and $\alpha=1$.
The red histograms and black solid lines show the results of Monte Carlo simulations 
and analytical solutions of Equations (\ref{eq:analytical}), respectively.
The thin and thick show distribution functions at $t/\tau_0=5\times10^6$ and $10^7$, respectively.
\label{fig:4}}
\end{figure}

In order to confirm analytical solutions presented in the previous section, 
we perform test particle Monte Carlo simulations. 
We here focus on static, statistically homogenous and isotropic 
incompressible turbulence, that is, 
the velocity field, $\delta u_i({\bf x})$, is divergence free.
Such a vector field is numerically constructed by a summation
of many transverse waves \citep{giacalone99}.  
Simulation particles are isotropically and elastically scattered 
in the local fluid frame and move in a straight line between each scattering.
The mean scattering time is given by $\tau=\tau_0(p/p_0)^{\alpha}$. 
We use $10^4$ simulation particles with the initial four momentum 
$p_0=10mc$ and $100$ transverse waves in order to construct velocity fields, 
where $m$ and $c$ are the particle mass and the speed of light.
The mean amplitude of velocity fluctuations is taken to be 
$\langle \delta u^2\rangle = (0.05c)^2$. 
We set the maximum wavenumber to be $\tau_0ck_{\max}=10^{-1/3}$ 
for the Kolmogorov spectrum. 

We first discuss results of Monte Carlo simulations for 
the momentum-independent scattering, that is, $\alpha=\beta=0$. 
Figure 1 shows the evolution of the mean four momentum for 
$\alpha=0$ and $\tau_0 c k_0 = 10^{-1}$. 
Particles are accelerated, and simulation results are in good agreement 
with analytical solutions of Equations (\ref{eq:dm}), (\ref{eq:dturbsu}) 
and (\ref{eq:mean}). 
By comparing the growth rate of the mean momentum of 
simulation particles with Equation (\ref{eq:mean}), we can obtain the 
momentum diffusion coefficient of Monte Carlo simulations. 

Figure 2 shows the wavenumber dependence of the momentum 
diffusion coefficient, $\tilde{D}_0=D_{\rm TSA}\tau_0/p^2$ for $\alpha=0$. 
Simulation results are in good agreement with Equations (\ref{eq:dm}) and 
(\ref{eq:dturbsu}) as long as $\tau_0 c k_0 < 1$, but simulation results for the 
monochromatic spectrum deviate from Equation (\ref{eq:dm}) at $\tau_0ck_0 > 1$.
As already mentioned in Section 2, this is because our treatment is not valid when 
the particle mean free path is larger than the turbulent scale.
Furthermore, we have confirmed that the Kolmogorov-like turbulent 
cascade (blue) enhances the turbulent shear acceleration.

Figure 3 shows the distribution function, $\tilde{p}dN/d\tilde{p}\propto p^3f(p,t)$, 
at $t/\tau_0=10^5$ for $\alpha=0$ and $\tau_0 c k_0 = 10^{-1}$. 
Simulation results (histograms) are in excellent agreement 
with analytical solutions of Equation (\ref{eq:analytical0}) (solid lines) 
for the monochromatic (red) and Kolmogorov (blue) spectra.

Figure 4 shows the distribution function at $t/\tau_0=5\times10^6$ and $10^7$ 
for the monochromatic spectrum with $\tau_0 c k_0 = 10^{-2}$ 
and the Bohm-like diffusion, that is, $\alpha=\beta=1$.
Simulation results (histograms) are in excellent agreement 
with analytical solutions of Equation (\ref{eq:analytical}) (solid lines) 
except for above $p/p_0\sim10^2$. 
As mentioned above, the disagreement is due to $\tau(p) v k_0 > 1$ 
at $p/p_0 >10^2$. 

Therefore, we have confirmed that the ensemble-averaged transport equation 
for incompressible turbulence describes the turbulent shear acceleration and, 
that is valid as long as the turbulent scale, $\sim k^{-1}$, is larger than the 
particle mean free path, $\tau v$.
 
%%%%%%%%%%%%%%%%%%%%%%%%%%%%%%%%%%%%%%%%%%%%%%%%%%%%%%%%%%%%%%%%%%%%%%
\section{Discussion}
We first discuss another important effect of turbulence on the particle transport.
\citet{bykov93} shows that turbulence enhances spatial diffusion. 
For strong turbulence, $\kappa_{\rm turb}$ becomes of the order of 
$L_0\sqrt{\langle \delta u^2\rangle}$  \citep{bykov93}, 
where $L_0$ is the injection lengthscale of 
turbulence, so that spatial diffusion of particles with a 
small mean free path is dominated by turbulent diffusion 
and an energy-independent diffusion is realized. 
The ratio of the turbulent diffusion and the Bohm diffusion, $\kappa_{\rm Bohm}$, 
is given by
\begin{equation}
\frac{\kappa_{\rm turb}}{\kappa_{\rm Bohm}} \sim 3\times10^6~
\left(\frac{p}{m_{\rm p}c}\right)^{-1}
\left(\frac{\sqrt{\langle \delta u^2\rangle}}{c}\right) 
\left(\frac{B}{1~{\rm \mu G}}\right) 
\left(\frac{L_0}{1~{\rm pc}}\right)~~,
\end{equation}
where $m_{\rm p}$ and $B$ are the proton mass and magnetic field, respectively. 
Therefore, turbulent diffusion of energetic particles could be important 
in SNRs, PWNe, astrophysical jets, etc.

From Equation (\ref{eq:dturbs}), the acceleration timescale, 
$t_{\rm acc}=p^2/D_{\rm TSA}$, of the turbulent shear acceleration 
for the Kolmogorov spectrum of the case $k_{\rm res}<k_{\rm max}$ 
is represented by 
\begin{eqnarray}
t_{\rm acc}&=&\frac{30}{\left\{\tau(p)vk_0\right\}^{2/3}} \frac{v^2}{\langle \delta u^2\rangle} \tau(p) \nonumber \\
&=&9\times10^6~{\rm sec} \nonumber \\
&&\times \left(\frac{p}{m_{\rm p}c}\right)^{1/3} 
\left(\frac{\langle \delta u^2\rangle}{c^2}\right)^{-1}
\left(\frac{B}{1~{\rm \mu G}}\right)^{-1/3} 
\left(\frac{L_0}{1~{\rm pc}}\right)^{2/3}~~,
\end{eqnarray}
where we have assumed $L_0=2\pi /k_0$ and the Bohm diffusion,  
$\tau(p)=p/(eB)$, in the last equation.
Therefore, particles can be accelerated to relativistic energies by 
large-scale turbulence in many astrophysical objects. 
In addition, if particles are initially accelerated at the shock, 
large-scale turbulence can change energy spectra of 
the accelerated particles in the shock downstream region. 

Next, we compare the turbulent shear acceleration and 
particles acceleration by small-scale incompressible turbulence, that is,  
the second order acceleration by Alfv{\'e}n waves \citep{skilling75}. 
The momentum diffusion coefficient of the second order acceleration 
by Alfv{\'e}n waves is given by $D_{A}\sim p^2v_{\rm A}^2/(9\kappa)$, 
where $v_{\rm A}$ is the Alfv{\'e}n velocity. 
The ratio of the turbulent shear acceleration and the second order acceleration 
by Alfv{\'e}n waves is given by
\begin{equation}
\frac{D_{\rm TSA}}{D_{\rm A}} \sim 
\frac{\langle\delta u^2\rangle}{v_{\rm A}^2}
\left(\frac{\tau v}{L_0}\right)^{2/3}~~,
\end{equation}
where we have adopted Equation (\ref{eq:dturbs}) as $D_{\rm TSA}$.
Therefore, the turbulent shear acceleration could be more efficient than 
the second order acceleration by Alfv{\'e}n waves for super-Alfv{\'e}nic turbulence 
($\sqrt{\langle\delta u^2\rangle} > v_A (\tau v / L_0)^{1/3}$) . 
In other words, the turbulent shear acceleration becomes important 
when there are strong magnetic field fluctuations ($\delta B/B_0 >1$) 
because the plasma velocity fluctuation by Alfv{\'e}n waves, $\delta u$, 
is represented by $\delta u = v_{\rm A} (\delta B/B_0)$, where $\delta B$ 
and $B_0$ are the fluctuated and mean magnetic fields, respectively.
Such a situation is expected to be realized in the downstream region of a high Alfv{\'e}n 
Mach number shock \citep{giacalone07,inoue09}.

We have considered only isotropic diffusion and nonrelativistic turbulence 
in this Letter. 
Spatial diffusion is generally anisotropic because of the magnetic field. 
Isotropic diffusion is realized 
when magnetic field fluctuations with lengthscale comparable to the particle mean free path are large ($\delta B/ B_0 >1$) \citep[e.g.][]{giacalone99}.
Therefore, as discussed above, the turbulent shear acceleration 
is important when isotropic diffusion is realized.
Simple extensions to anisotropic diffusion and relativistic 
turbulence are straightforward because extensions of Equation 
(\ref{eq:transporteq}) have already been provided by \citet{webb89,williams93}. 
This calculation will be addressed in the future work.

%%%%%%%%%%%%%%%%%%%%%%%%%%%%%%%%%%%%%%%%%%%%%%%%%%%%%%%%%%%%%%%%%%%%%%
\section{Summary}
In this Letter, we have derived a particle transport equation 
averaged over random plasma flows in order to understand 
particle acceleration in incompressible turbulence with a larger 
lengthscale than the particle mean free path. 
We have considered ensemble average of the extended transport equation 
provided by \citet{webb89,williams93}.
This is a simple extension of previous work that considered ensemble 
average of the transport equation provided by \citet{parker65}.
We have found that the turbulent shear acceleration by incompressible 
turbulence becomes important in small scale for Kolmogorov-like turbulence.
Moreover, we have performed Monte Carlo simulations and confirmed 
the turbulent shear acceleration. 
Recent simulations show that turbulence is produced in many 
astrophysical objects, so that turbulent diffusion and turbulent 
acceleration are expected to be important.

\acknowledgments
We thank T. Inoue and R. Yamazaki for useful comments about turbulence 
and simulation.
This work is supported in part by grant-in-aid from the Ministry of Education, 
Culture, Sports, Science, and Technology (MEXT) of Japan, No.~24$\cdot$8344.

\end{document}